\documentclass[12pt]{article}
\usepackage{amssymb}
\def\inh{\vskip 0.075truein \noindent\hangindent=12 pt \hangafter=1}

\usepackage{amsthm}\theoremstyle{remark}
\usepackage{graphicx}

\newcommand{\bte}{\begin{quote}\begin{theorem}}
\newcommand{\ete}[1]{\label{#1}\end{theorem}\end{quote}}
\newcommand{\bcom}{\begin{quote}\end{quote}}
\newcommand{\bex}{\begin{quote}\begin{example}}
\newcommand{\eex}[1]{\label{#1}\end{example}\end{quote}}
\newcommand{\bcon}{\begin{quote}\begin{conclusion}}
\newcommand{\econ}[1]{\label{#1}\end{conclusion}\end{quote}}
\newcommand{\bdefi}{\begin{quote}\begin{definition}}
\newcommand{\edefi}[1]{\label{#1}\end{definition}\end{quote}}

\newcommand{\blem}{\begin{quote}\begin{lemma}}
\newcommand{\elem}[1]{\label{#1}\end{lemma}\end{quote}}

\newcommand{\bpr}{\begin{quote}\begin{problem}}
\newcommand{\epr}[1]{\label{#1}\end{problem}\end{quote}}

\newcommand{\f}{\frac}
\newcommand{\p}{\partial}
\newcommand{\n}{\nonumber \\}

\newcommand{\beq}{\begin{eqnarray}}
\newcommand{\eeq}[1]{\label{#1}\end{eqnarray}}
\newcommand\eq[1]{(\ref{#1})}
\newcommand{\bfi}{\begin{figure}[24]}
\newcommand{\efi}[1]{\caption{\label{#1}}\end{figure}}

\newcommand{\res}{respectively}
\newcommand\gl{\left}
\newcommand\gr{\right}
\newcommand{\bfm}[1]{\mbox{\boldmath ${#1}$}}

\newcommand{\CA}{{\cal A}}

\newcommand{\CE}{{\cal E}}

\newcommand{\CK}{{\cal K}}

\newcommand{\Ga}{\alpha}

\newcommand{\Gl}{\lambda}

\newcommand{\Gr}{\varrho}

\newcommand{\Go}{\omega}

\newcommand{\Gz}{\zeta}

\newcommand{\az}[1]{Sect.$\!$ \ref{#1}}
\newcommand\D{\,\mathrm{d}}

\newcommand{\bexe}{\begin{quote}\begin{exercise}\inh}
\newcommand{\eexe}[1]{\label{#1}\end{exercise}\end{quote}}

\usepackage{graphics,graphicx}
\usepackage{color}
\begin{document}
{\large
\title{Mechanical wave momentum\\
 from the first principles}
\author{Leonid I. Slepyan}
\date{\small{{\em School of Mechanical Engineering, Tel Aviv University\\
P.O. Box 39040, Ramat Aviv 69978 Tel Aviv, Israel}}\\
 }}

\maketitle

\vspace{0mm}\noindent
{\bf Abstract}
\noindent
Axial momentum carrying by waves in a uniform waveguide is considered based on the conservation laws and a kind of the causality principle. Specifically, we examine (without resorting to constitutive data) steady-state waves of an arbitrary shape, periodic waves which speed differs from the speed of its form and binary waves carrying self-equilibrated momentum.  The approach allows us to represent momentum as a product of the {\em wave mass} and the wave speed. The propagating wave mass, positive or negative, is the excess of that in the wave over its initial value. This general representation is valid for mechanical waves of arbitrary nature and intensity. The finite-amplitude longitudinal and periodic transverse waves are examined in more detail. It is shown in particular, that the transverse excitation of a string or an elastic beam results in the binary wave. The closed-form expressions for the drift in these waves functionally reduce to the Stokes' drift in surface water waves (a half the latter by the value). Besides, based on the general representation an energy-momentum relation is discussed and the physical meaning of the so-called ``wave momentum" is clarified.

\vspace*{10mm}\noindent

\vspace{10mm}\noindent
Corresponding author (Leonid Slepyan), email: leonid@eng.tau.ac.il

\section{Introduction}
Along with energy, momentum plays a defining role in wave actions, and different related questions were debated since Lord Rayleigh's works on the theory (Rayleigh, 1902, 1905). Brillouin (1925), McIntyre (1981), Ostrovsky and Potapov (1988), Peskin (2010), Falkovich (2011) and Maugin and Rousseau (2015) are among others who considered various aspects of this topic (a comprehensive list of references can be found in the latter book by Maugin and Rousseau ).

Note, however, that momentum, in contrast to the energy, generally is not a nonnegative value and it is very sensitive to the formulation. In particular, it may be lost with linearization. By definition, momentum is the product of density and the particle velocity, $p=\Gr \bfm{v}$, and (possibly small) variations of both of them should be taken into account. The question is whether it is possible to represent the quantity in a classical manner as the product of a mass and the wave velocity, where only one of the multipliers, the mass can be variable. Below, we introduce such a representation.

Most related results obtained correspond to specific problems and low-intensity waves. The question is whether a straightforward way exist for determining the momentum of high-intensity waves. We show that the above-discussed representation lets do it.

In dispersive sinusoidal waves, the energy propagates with the group velocity. Does the latter play a similar role for momentum? We show that affirmative reply is correct.

Under a linearized formulation where momentum is lost, could it possible to extract it based on the simplified description of the wave?
We show that such a possibility does exist, and the lost momentum can be obtained considering the mass distribution or even only the waveform (as, for example, in the case of the flexural wave considered below).

In the wave theory, along with the classical definition of momentum, the so-called ``wave momentum" $-\Gr_0u'v$ is considered ($\Gr_0 $ is the initial density, $u'$ and $v$ are the derivative of the displacement and the particle speed).  It is ``...  a much debated notion" (Maugin and Rousseau, 2015, p. 2). ``Of course, considering a ``definition" of wave momentum ... would be quite reasonable from a strict mathematical viewpoint, but the physical meaning would be doubtful." (ibid, p.14). Below, based on the general representation the physical meaning of the "wave momentum" is given.

In the present work, we show that the essence of the issue and useful relations immediately follow from the mass and momentum conservation with the assumption that ahead of the wave the waveguide is in a uniform static state until the wave arrives (this is a kind of the {\em causality principle}). Abstracting from constitutive relations and other particularities a greater clarity can be retrieved, and the results obtained directly from these first principles are valid for mechanical waves of any nature, form and intensity.

We consider a homogeneous straight-line waveguide of an arbitrary cross-section and three types of mechanical waves: a steady-state wave of an arbitrary shape, a periodic wave which speed differs from the speed of its form (for a linear sinusoidal wave these are the group and phase velocities, \res) and a binary wave carrying a self-equilibrated momentum.

The momentum follows as a product of the {\em wave mass} and the wave speed. Based on the general representation the wave-induced drift and the connection between the momentum and energy are discussed. Periodic longitudinal and flexural waves are considered as nontrivial examples.

Along with the {\em reference frame} associated with this initial state, we use a frame moving with the wave. Eulerian (spatial) variables are used, unless otherwise is specified. We denote the longitudinal coordinate by $x$. The mass and momentum densities per unit $x$-length and the $x$-component of the particle velocity are denoted by $m=m(x,t)$, $p=p(x,t)$ and $v= v(x,t)$, \res. In the initial state of the waveguide, ahead of the wave, $m= m_0, v=0$ (or asymptotically equal to these quantities as $x\to\infty$). As usual, the prime and dot mean the partial derivatives on the coordinate $x$ and time $t$.

\section{Steady-state wave}
We are interested in axial momentum of a wave of a general form and intensity propagating along the $x$-axis with the speed $c$ (note that the wave can possess both the axial and angular momenta, see, e.g., Slepyan at al, 1995). In the frame moving with the wave, the mass distribution is assumed to be fixed over the wave domain as well as ahead of the wave where the uniform waveguide is in the initial static state (condition $\CA$).

\subsection{The wave mass and momentum}
As the first state consider the wave in the moving frame, where the waveguide medium flows through the wave domain, and momentum in front of the wave is equal to $- m_0 c$.
For such a steady movement, it follows from the mass conservation and the above condition $\CA$ that
\beq \f{\p p}{\p x} =\f{\p}{\p x} \int_S \Gr v \D S = - \f{\p m}{\p t}=0 ~~~\Longrightarrow ~~~p = p_- = - m_0 c\,,\eeq{nn1}
where $\Gr$ and $v$ are density and the $x$-component of the particle velocity, and $S$ is the cross-section area. Now, adding the corresponding right-directed rigid momentum $p_+ = m c$ (to return to the reference frame) we obtain the momentum density
\beq p =m_w c\,, ~~~m_w= m -m_0\,.\eeq{ssw2}
The excess of the mass in the wave $m_w$ can be called the wave mass (per unit length). It does not consist of the same particles but propagates as the wave. Note that both these parameters, the wave mass and momentum, can be positive, zero or negative. In the latter case, the momentum directs opposite to the wave.

A step wave is the simplest example. The mass conservation reads as
\beq \Gr(c-v) = \Gr_0 c~~~\Longrightarrow~~~p = \Gr v = (\Gr-\Gr_0)c\,.\eeq{ssw3}
Recall that $v$ is the particle velocity and $\Gr$ and $\Gr_0$ are the actual and initial mass densities.

The advantage of the representation \eq{ssw2} as compared with the initial definition, $p=mv$, is that it contains only one variable $m$. In certain cases, this allows immediately indicate the presence of momentum and to determine it based on the linearized expression of the wave, as in the case of a flexural wave considered below.

\subsection{The momentum, energy and wave shift}
Assuming the flow in the first state to be irrotational consider a stream filament. Let $s, v_s $ and $\Gr_s$ be the curvilinear coordinate, the speed along it and the mass per unit $s$-length, \res. For brief, we now take the case where $s(x)$ is a single-valued function. In a similar manner as above, we call $\Gr$ and $\Gr_0$ the densities of the mass per unit $x$-length in the actual and initial states of the filament. Now $v$ is the $x$-projection of the velocity $v_s$, and the mass conservation reads as $\Gr_s v_s = - c\Gr_0$. It is also convenient to introduce the stretch $\Gl$ of the distance between (close) filament {\em material} cross-sections normal to the $x$-axis, $\Gl=\Gr_0/\Gr$.

The expressions \eq{nn1} and \eq{ssw2} are also valid for the filament. Indeed
\beq \Gr= \Gr_s \D s/\D x\,,~~~v_- = v_s\D x/\D s\,,~~~\Longrightarrow~~~\Gr v_- = \Gr_s v_s=-\Gr_0 c\,, \eeq{mes1}
where $v_-$ is the particle speed in the first state, and
\beq p_- = \Gr v_- =-\Gr_0 c\,,~~~p = p_-+\Gr c =\Gr_w c\,,~~~\Gr_w= \Gr - \Gr_0\,.\eeq{mes2}

The filament wave mass density (per unit $x$-length), $\Gr_w$, in Eulerian (E) and Lagrangian (L) description is
\beq \Gr_w = \left\{ \begin{array}{ll}
  \Gr_0 (1/\Gl - 1) = - \Gr_0u'~~~& (E) \\
  \Gr_0(1-\Gl) = - \Gr_0u'& (L)\,,
 \end{array}\right.\eeq{wmes1}
where $u$ is the longitudinal displacement (recall that here $\Gl$ is the stretch of the distance but not of the medium element). The value of $\Gl$ is the same for both expressions, whereas $u'$ is different if $\Gl\ne 1$.

The kinetic energy density is
\beq \CK = \f{\Gr}{2}[(c+v_s\cos\Ga)^2 + (v_s\sin\Ga)^2]= \f{\Gr c^2}{2}\gl[1+\gl(\f{\Gr_0}{\Gr_s}\gr)^2-2\f{\Gr_0}{\Gr}\gr],\eeq{tem1}
where $\Ga$ is the angle between the filament and the $x$-axis.
In particular, if $\Gr_s=\Gr_0$ this expression becomes
\beq  \CK = p c = \Gr_w c^2\,.\eeq{tem2}
Note that the kinetic$-$potential energy partition is considered in Slepyan (2015).

An arbitrary wave in a string under a constant tensile force $T$ is an example. The wave propagates with the speed $c=\sqrt{T/\Gr_0}$ (the string is in equilibrium while moving uniformly with this speed along any fixes trajectory). The above relations are valid for this case, and the wave mass corresponding to an $x$-segment is $\Gr_0(\ell_s-\ell)$, where $\ell_s$ is the arc length and $\ell$ is the segment length.

The wave passing can be accompanied by a progressive shift. In the stream filament, there is a one-to-one correspondence between the shift and momentum density. Indeed, the shift per unit $x$-length
\beq  q = \f{1-\Gl}{\Gl}= -u' ~~(\mbox{E})\,,~~~q= 1-\Gl =- u'~~(\mbox{L})~~\Longrightarrow~~q=\f{p}{\Gr_0 c}~~(\mbox{E and L})\,.\eeq{tws1}
The wave shift is typical for solitary waves. The well-known Stokes' drift in surface water waves (Stokes, 1847) is also an example.

\subsection{Physical meaning of the ``wave momentum"}
The so-called "wave momentum", $-\Gr_0u'v$, can be treated as the momentum without a zero-mean term. Indeed, represent the momentum density with a linear part separated. With refer to \eq{wmes1} we obtain
\beq p = \Gr v = p_0 + p_w \,,~~~p_0= \Gr_0 v\,,~~~ p_w = (\Gr - \Gr_0) v=-\Gr_0 u' v\,,\eeq{wmm1}
where $\Gr - \Gr_0$ is the {\em wave mass density}.
The separation makes sense for the case of a periodic wave with a zero mean $v$. Here the first term vanishes on the averaged, and only the second one remains. Thus, in this case, the "wave momentum" coincides on the average with the original definition of this quantity.

\section{Binary wave}\label{bw}
A binary wave in a helical fiber was described by Krylov and Slepyan (1997). Caused by an axial force with no torque it was a wave consisting of two parts rotating with opposite directed angular momenta of the same value. Similarly, a binary wave with a self-equilibrated axial momentum can be emitted without axial action.

Consider the corresponding piecewise constant two-step wave consisting of a forerunner (the front part of the wave) and the successive part (the wave excited transversely). Let the former be a longitudinal wave propagating with the speed $c_+$,  the mass density $m_+$ and particle velocity $v_+$, whereas $m$ and $v$ be the corresponding values for the flexural wave, and let $D$ be the speed of the interface between these parts.

The mass conservation relations are
\beq (c_+-v_+)m_+ = c_+m_0\,,~~~(D- v)m =(D-v_+)m_+\,.\eeq{bw1}
The axial momentum per unit time in the flexural wave, $P$, and in the forerunner, $P_+$, can be obtained in a similar way as above for the single wave. However, the fact should be taken into account that in front of the former, the mass density and speed are $m_+$ and $v_+$, not $m_0, c$. In this problem, the energy densities $\CE$ and $CE_+$ as well as the energy flux speeds $c_g$ and $c_+$ (for the flexural wave and forerunner, \res) are to be taken into account. Note that, in a true steady-state wave, $c_g=c$; however, for greater generality, we denote them as the group and phase speeds, \res. In a linear longitudinal wave, where the kinetic and potential energies are equal, $\CE_+= m_+v_+^2$.

With refer to the mass conservation relations \eq{bw1} the momentum densities per unit time can be expressed as
\beq P = m v(D-v)= [(m-m_+)D + (m_+-m_0)c_+](D-v)\,,~\n P_+=(m_+-m_0)c_+(c_+ -D) \,.\eeq{nn2}
For the four unknowns, $v, v_+, m_+$ and $D$ ($c_g, c_+, m_0$ and $m=m_0+m_w$ are the input data) there are four equations, two in \eq{bw1} plus the momentum and energy conservation relations
\beq  P+P_+=0\,,~~\CE(c_g-D) = \CE_+(c_+-D)\,.\eeq{nn3}
An explicit solution can be presented for a low-intensity wave where $|m_w/m_0|\ll 1$. The corresponding asymptotic values are
\beq v\sim \f{c_g c_+}{c_++c_g}\f{m_w}{m_0}\,,~~~v_+ \sim -\f{c_+c_g^2}{c_+^2-c_g^2}\f{m_w}{m_0}\,,\n
m_+ -m_0 \sim - \f{c_g^2}{c_+^2-c_g^2}m_w\,,~~~
c_g -D \sim (c_+-c_g)\f{ m_0c_+^2c_g^4}{(c_+^2-c_g^2)^2\CE}\gl(\f{m_w}{m_0}\gr)^2.\eeq{arbw}

The momentum densities per unit time follow as
\beq P=-P_+ = mv(D-v) \sim m_0c_g v \sim \f{c_+c_g^2 m_w}{c_++c_g}\,.\eeq{mbwf}
Thus, the densities per unit length asymptotically equal to
\beq p\sim \f{P}{c_g} \sim \f{c_+c_g m_w}{c_++c_g}\,,~~~p_+ \sim - \f{P}{c_+-c_g}\sim -  \f{c_+c_g^2 m_w}{c_+^2-c_g^2}.\eeq{mbwf2}
In addition, we compare the energy densities of these parts of the binary wave. For the linearised formulation the potential and kinetic energies should be equal. It follows that
\beq \CE = m v^2\sim \f{m_0 c_g^2c_+^2}{(c_++c_g)^2}\f{m_w^2}{m_0^2}\,,~~~
\CE_+ = m_+ v_+^2\sim \f{m_0 c_g^4c_+^2}{(c_+^2-c_g^2)^2}\f{m_w^2}{m_0^2}\,,\eeq{123}
and the ratio is
\beq \f{\CE_+}{\CE} \sim \f{c_g^2}{(c_+-c_g)^2}\,.\eeq{124}

The binary wave arising under transverse harmonic excitation of an elastic beam is considered in \az{lin}.

\section{Periodic waves}
Here we consider a periodic wave which speed $c_g$ differs from the speed $c$ of its form. Usually, such a wave is characterized by a quasi-front which is considerable in a slowly expanding transition area in a vicinity of $x=c_gt$. In the present study, taking into account the fact that as the wave propagates the quasi-front area becomes asymptotically negligible, we consider it as a periodic wave with the front at $x=c_g t$.

We use the $c_g$-moving frame where the wave does not propagate but the waveform moves with the speed $c-c_g$. In this state, there is not a fixed but a periodic distribution of the mass and particle velocities, and the expressions for for the mass conservation and for $p_-$ \eq{nn1} and $p_+$ \eq{ssw2} are still valid being averaged over the period. We have
\beq \langle p_-\rangle = -m_0 c_g\,,~~~\langle p_+\rangle = \langle m\rangle c_g\,,~~~\langle p\rangle = m_w c_g\,,~~~m_w= \langle m -m_0 \rangle\,.\eeq{pw1}
So, momentum propagates with the same speed $c_g$ as the energy.

The expressions in \eq{wmes1} can also be used for the filament
\beq \langle p\rangle =-\Gr_0\langle u' \rangle c_g = \f{\Gr_0 c_g}{L}[\langle u(x)-u(x+L) \rangle]\,,\eeq{pw2}
where the value of the period $L$ depends on what the formulation, Eulerian or Lagrangian, is used.

The period-averaged drift speed is
\beq v_d =-\langle u'\rangle c_g =\f{\langle p\rangle}{\Gr_0}\,.\eeq{pwd}

\section{Longitudinal and flexural waves}

\subsection{Longitudinal oscillating wave}
Here we consider a one-dimensional, longitudinal, periodic wave of a zero mean particle velocity. The latter condition can be formulated concerning the velocity of fixed particles (as the wave excited by the {\em Lagrangian emitter}) or regarding the particle velocity detected at a fixed spatial point (the wave excited by the {\em Eulerian emitter}). The results corresponding to these different formulations could be different (see Falkovich, 2011, pp. 75-76). In the former case considered by McIntyre (1981) momentum is at zero regardless of the wave form and intensity. The latter case was examined by iterations for a small ratio of the particle displacements to the wavelength (Falkovich, 2011). Below, based on the representation presented in \eq{mes2} and \eq{wmes1} we express momentum in a closed form valid for the oscillating wave of a finite amplitude. The only condition is assumed that the particle velocity is below the wave speed.

Consider a 1D longitudinal wave where the particle velocity can be represented in the form
\beq v=\f{\D x(t)}{\D t}=v_0 f\gl(t-\f{x(t)}{c}\gr)H(ct-x(t))\,,\eeq{lw1}
where $-1\le f\le 1$, $0< v_0<c$ and for any $x=x_*$ fixed and period $T$
\beq \int_T f(t-x_*/c)\D t=0\,.\eeq{lw2}
The displacement follows from this as
\beq  u= v_0\int_{x/c}^t f\gl(t-\f{x(t)}{c}\gr)\D t = v_0 c\int_0^\Gz \f{f(\Gz)\D \Gz}{c-v_0 f(\Gz)}\,,~~~
\Gz = t-\f{x(t)}{c}\,.\eeq{lw3}
Thus
\beq u' = - \f{v_0 f(\Gz)}{c-v_0f(\Gz)}\eeq{lw4}
and
\beq p = \f{\Gr_0 v_0 c f(\Gz)}{c-v_0f(\Gz)}\eeq{lw5}
In particular, if $v\ll c$
\beq p\sim \Gr_0 v_0 f(\Gz) + \f{\Gr_0 v_0^2}{c}f^2(\Gz) + ...\,, \eeq{lw6}
and if $f(\Gz) = \sin (\Go t - k x)$ the averaged wave drift and momentum are
\beq v_d =- c u' \sim \f{v_0^2}{2c} = \f{a^2\Go^2}{2c}\,,~~~\langle p\rangle =\Gr_0 v_d\,,\eeq{lw7}
where $a=v_0\Go$ is the wave amplitude. Note that for a dispersive wave the phase speed $c=\Go/k$ must be replaced by group speed $c_g$. In the case $c=c_g$
\beq v_d \sim \f{1}{2}a^2\Go k\,,~~~\langle p\rangle =\Gr_0 v_d\,.\eeq{lw8}
Note that the drift speed induced by the 1D low-intensity longitudinal sinusoidal wave, that reflected by the expression in \eq{lw8}, is functionally the same but twice as less as the Stokes' drift on the surface of a 2D wave propagating on deep water (see, e.g., Lighthill, 1978). In the next section, we show that the momentum of the flexural sinusoidal wave appears the same as for the longitudinal one.

Also, note that the above results do not follow directly from the linearized formulation, which is incorrect for any value turned out to be zero. So, if the linearized description evidences that the wave does not possess momentum, it does not mean that the latter is absent. It should be added that the linearized mathematical theory can be perfect by itself but does not necessarily reflect the corresponding physical applications correctly. In such a case, further analysis as done above is required.
\subsection{Flexural-longitudinal binary wave}\label{lin}
Consider a semi-infinite string or an elastic beam, $0<x<\infty$. Let a flexural wave, be exited by a transverse periodic action at $x=0$, and the transverse displacements in the wave area be
\beq w = af(x- ct)\,.\eeq{flw1}
Here the representation of momentum through the wave mass is especially useful.
Indeed, since no axial force is applied, we may conclude that in the established regime, the string axis is not stretched, and the wave mass is
\beq m_w = m_0\sqrt{(1+u')^2+(w')^2}\,,\eeq{flw2}
and its value averaged over the period $\ell$ is
\beq \langle m_w\rangle =\f{ 1}{\ell}\int_\ell m_w \D x = \f{L}{\ell}-1\,,\eeq{flw3}
where $L$ is the arc length corresponding to the hord length $\ell$.

Thus, the axial momentum appeared without axial action, which may happen if the flexural wave excites the longitudinal one with the same but opposite directed momentum. That is, the transverse action results in a binary wave as discussed above.

As an example consider a sinusoidal flexural wave in a linearly elastic beam. For a large wave-length-to-amplitude ratio, the classical Bernoulli-Euler dynamic equation is valid, possibly with a term corresponding to the tensile force. The equation is written regarding the transverse displacements, and it may seem that there is no axial momentum; however, it is not so as we already know.

To determine the momentum of the binary wave, using the results obtained in \az{bw}, we need the input parameters, densities of the initial mass as well as the averaged wave mass and energy for the flexural wave, and also the wave speeds, $c_g$ and $c_+$.

In terms of the transverse and longitudinal displacements, $w$ and $u$, \res, with  small derivatives $w'$ and $u'$, the binary wave can be described as follows. Since there is no axial tensile force in the flexural wave region, we based on the equation
\beq c_+^2r^2w^{IV}+ \ddot{w}=0\,,\eeq{twe1}
where $r=\sqrt{J/A}$ is the radius of inertia moment $J$ of the beam cross-section area $A$. The flexural wave is
\beq w= a \cos(\Go t - k x)\,,~~~k = \sqrt{\f{\Go}{c_+r}}.\eeq{wnfw1}
The frequency as a function of $k$ and the group speed $c_g$ are
\beq \Go =c_+r k^2\,,~~~c_g = 2c_+r k=2c_+\sqrt{\f{\Go r}{c_+}}\,.\eeq{wnfw2}
The initial and actual mass densities can be expressed as
\beq m_0= A\Gr_0\,,~~~m = \f{m_0}{\Gl}\,,~~~\Gl =\f{1}{1-u'}. \eeq{33}
The derivative $u'$ can be found from the fact that the arc length of a beam material segment remains the same as its corresponding initial length. This results in the equality
\beq  \sqrt{1+(w')^2} =1 - u' ~~~\Longrightarrow~~~u'=1- \sqrt{1+(w')^2} \sim - \f{(w')^2}{2}\,.\eeq{44}
Thus, the mass and the wave mass are
\beq m = m_0(1-u')\sim  m_0\gl(1+\f{(w')^2}{2}\gr)\,,~~~m_w=m-m_0\sim m_0\f{(w')^2}{2}\,,\n
\langle m_w\rangle = \f{m_0 a^2k^2}{4}= \f{m_0 a^2 \Go}{4c_+r}\,.\eeq{55}
Finally, the flexural wave energy density is
\beq \CE = \Gr_0(\dot{w})^2= \Gr_0 a^2 \Go^2\sin^2(\Go t - k x)\,,~~~\langle \CE\rangle =\f{1}{2}\Gr_0 a^2 \Go^2\,.\eeq{66}

Now all the values presented in \eq{arbw} - \eq{mbwf2}, except the expression for $c_g-D$, can be calculated replacing the variable values by their averaged representations. As to the $c_g-D$ difference, we note that the original equation (the second relation in  \eq{bw1}) contains not negligible products of variable values, whereas the product averaged does not equal to the product of the averaged factors. However, for the low-intensity wave considered here, the difference is very small; it is of the order of $(ak)^4$. So, with the accuracy adopted in the linearized formulation, it only evidences that $D<c_G$ with $D\sim c_g$.

In particular, the  ratio of the energy densities defined in \eq{124} now is
\beq \f{\langle \CE_+\rangle}{\langle \CE\rangle} \sim \f{4r^2k^2}{(1- 2rk)^2}\sim 4r^2k^2 \ll 1\,.\eeq{77}
since $|rk| \ll 1$ by the definition. This estimation of the energy ratio is also valid for the total energies. Indeed
\beq \f{\CE_+ (c_+-D)}{\CE (D-v)} \sim  \f{\CE_+ c_+}{\CE c_g}\sim 2rk\,.\eeq{88}
Nevertheless, the longitudinal wave possesses the same momentum, by its value, as the much more powerful flexural wave. However, the momentum is rather small. It asymptotic value coincides with that following directly from the mass conservation \eq{ssw2}
\beq P =-P_+= m_+v_+(c_+-D) \sim m_0c_+v_+\sim m_w c_g^2\,,~~~p = m_w c_g\,.\eeq{fefm}
So, the momentum carrying by the flexural wave can be found using the expression in \eq{ssw2}, with $c=c_g$, disregarding the forerunner. Surprisingly, it is asymptotically the same as in the longitudinal wave. Namely, as follows from \eq{55} with the expression for the group speed in the beam, $c_g =2\Go/k$, the averaged value of momentum is
\beq \langle p\rangle \sim \f{m_0 a^2k^2c_g}{4}= \f{1}{2} m_0 a^2 \Go k\,,\eeq{ad1}
and the drift is also a half of that in the surface of the water waves
\beq v_d = \f{1}{2}a^2 \Go k\,.\eeq{ad2}

Thus, in this case, the flexural wave carries a momentum, while the associated longitudinal extensional wave has equal, opposite directed momentum. At the same time, in such a problem but for the infinite beam, due to the symmetry, there is no momentum in the flexural waves, while the associated longitudinal waves possess momenta directed to the origin.

\section{Conclusions}
Based on the mass conservation and the causality principle the {\bf axial momentum} density per unit length (or its period-averaged value) is represented by product, $p=m_w c$, of the wave the {\em wave mass} and speed (the energy flux speed). The latter is the (positive or negative) excess over its initial value. The wave mass does not consist of the same particles but propagates together with the wave energy transmitting through the waveguide. In the case of inextensibility, the wave energy $\CE = m_w c^2$.

The representation is valid for {\bf steady-state, periodic} and {\bf binary} mechanical waves of any nature, form and intensity.

Momentum propagates with the {\bf wave speed} the same as the energy (in the case of a linear sinusoidal wave, it is the group speed).

A physical meaning of the so-called {\bf wave momentum}, $-\Gr_0u'v$, is clarified for a periodic wave with a zero mean particle speed $v$. It is the momentum without zero mean part.

Explicit expressions for the wave drift and momentum are presented for an arbitrary {\bf longitudinal wave} of a finite amplitude.

It is shown that momentum does exist in {\bf transverse waves}. Under the transverse excitation of a string or a beam, the flexural-longitudinal binary waveforms carrying self-equilibrated axial momentum. The longitudinal part of the wave, excited in front of the flexural one, has a relatively low density in a relatively large region. Parameters of the wave are presented (explicitly for a low-intensity wave).

As a rule, {\bf linearization} eliminates the distinction between initial and current densities as well as between the Lagrangian and Eulerian formulations, the distinctions which should be taken into account in the momentum determination. Also, some needed values are reduced to zero. The above considerations suggest that the approximate results can be used but possibly with some additional data. In particular, the existence of momentum depends on the emitter type for a longitudinal sinusoidal wave, and on the boundary conditions related to the axial motion, in the case of the a flexural wave. In other words, if the momentum based on linearized formulation is zero, to find its value the formulation should be properly expanded.

Finally, concerning the {\bf forces} with which a wave acts on an object, we note that irrespective of momentum, there exists an energy release from the wave at a moving obstacle, and the corresponding configurational force remains nonzero at a zero limit of the speed. Also, the wave reflection results in a (Newtonian) force corresponding to the change of the momentum, if it exists. Both these factors take part in the action of a wave on the obstacle.

\vspace{10mm}
\vskip 18pt
\begin{center}
{\bf  References}
\end{center}
\vskip 3pt

\inh D.G. Andrews and M.E. McIntyre (1978). "An exact theory of nonlinear waves on a Lagrangian mean flow". Journal of Fluid Mechanics 89 (4): 609–646. Bibcode:1978JFM....89..609A. doi:10.1017/S0022112078002773

\inh Rayleigh, Lord, 1902. On the pressure of vibrations. Phyl. Mag. 3, 338-346.

\inh Rayleigh, Lord, 1905. On the momentum and pressure of gaseous vibrations, and on the connection with the virial theorem. Phyl. Mag. 10, 364-374.

\inh Brillouin, L., 1925.  Sur les tensions de radiation. Ann. Phys. 4, 528–586.

\inh Falkovich, G., 2011. Fluid Mechanics (A short course for physicists). Cambridge University Press. ISBN 978-1-107-00575-4

\inh Krylov, V. and Slepyan, L., 1997. Binary wave in a helical fiber. Physical Review B 55(21) (June 1),14067-14070.

\inh Lighthill J., 1978. Waves in fluids. Cambridge University Press, Cambridge, London, New York, Melbourne.

\inh Maugin, G.A., and Rousseau, M., 2015. Wave Momentum and Quasi-Particles in Physical Acoustics. World Scientific Series on Nonlinear 
Science Series A: Volume 88.

\inh McIntyre, M.E., 1981. On the “wave momentum” myth, J. Fluid Mech. 106, 331–347.

\inh Ostrovsky, A.A. and Potapov, A.I., 1988. The modulated waves in linear media with dispersion. The Nighny Novgorod (Gor'ky) University (in Russian).

\inh Peskin, C.S., 2010.  Wave momentum.  Courant Institute of Mathematical Sciences, New York University,\\
http://silverdialogues.fas.nyu.edu/docs/IO/24452/peskin.pdf

\inh Rayleigh, Lord, 1902. On the pressure of vibrations. Phyl. Mag. 3, 338-346.

\inh Rayleigh, Lord, 1905. On the momentum and pressure of gaseous vibrations, and on the connection with the virial theorem. Phyl. Mag. 10, 364-374.

\inh Slepyan, L., Krylov, V. and Parnes, R., 1995. Solitary Waves in
an Inextensible, Flexible, Helicoidal Fiber. Physical Review
Letters, 74, No. 14, 2725-2728.

 \inh Slepyan, L.I., 2015. On the energy partition in oscillations and waves.  Proc. R. Soc. A 471: 20140838.\\
DOI: 10.1098/rspa.2014.0838

\inh Stokes, G. G. 1847. On the theory of oscillatory waves. Trans. Camb. Phil. Soc. 8: 441–455.

\end{document}